\documentclass{article}
\usepackage{graphicx}

\evensidemargin=\oddsidemargin
\def\Title#1#2#3{%
    \baselineskip=18pt
    \begin{center}
          {\large\bf{#1} \\ }
          \bigskip\bigskip
          {#2} \\
          {#3} \\
    \end{center}}
\long\def\Abstract#1{%
         \bigskip
         \parbox{0.93\textwidth}{%
                 \begin{center}
                       {\bf Abstract} \\
                 \end{center}
                 \medskip{\baselineskip=14pt #1}
                 \vss}
         \bigskip}

\makeatletter
\renewcommand{\section}%
 {\@startsection{section}{1}{0pt}%
  {-3.25ex plus -1ex minus -.2ex}{1.5ex plus .2ex}%
  {\vspace*{5mm}\raggedright\large\bf }}
\renewcommand{\subsection}%
 {\@startsection{subsection}{2}{0pt}%
  {-2.25ex plus -.5ex minus -.2ex}{-1.5ex plus -.2ex}{\bf }}
\renewcommand{\subsubsection}%
 {\@startsection{subsubsection}{3}{0pt}%
  {-1.25ex plus -.2ex minus -.1ex}{-1.2ex plus -.2ex}{\bf }}

\makeatother

\begin{document}

\Title{The semiclassical limit of quantum gravity and the problem of time}%
{R. I. Ayala O\~na\footnote{E-mail: {\tt ayyala@sfedu.ru}},
M. B. Kalmykov\footnote{E-mail: {\tt mkalmykov@sfedu.ru}},
D. P. Kislyakova\footnote{E-mail: {\tt dkislyakova@sfedu.ru}} and
T. P. Shestakova\footnote{E-mail: {\tt shestakova@sfedu.ru}}}%
{Department of Theoretical and Computational Physics,
Southern Federal University,\\
Sorge St. 5, Rostov-on-Don 344090, Russia}

\Abstract{The question about the appearance of time in the semiclassical limit of quantum gravity continues to be discussed in the literature. It is believed that a temporal Schr\"odinger equation for matter fields on the background of a classical gravitational field must be true. To obtain this equation, the Born -- Oppenheimer approximation for gravity is used. However, the origin of time in this equation is different in works of various authors. For example, in the papers of Kiefer and his collaborators, time is a parameter along a classical trajectory of gravitational field; in the works of Montani and his collaborators the origin of time is introducing the Kucha\v r -- Torre reference fluid; in the extended phase space approach the origin of time is the consequence of existing of the observer in a fixed reference frame. We discuss and compare these approaches. To make the calculations transparent, we illustrate them with a model of a closed isotropic universe. In each approach, one obtains some Schr\"odinger equation for matter fields with quantum gravitational corrections, but the form of the equation and the corrections depend on additional assumptions which are rather arbitrary. None of the approaches can explain how time had appeared in the Early Universe, since it is supposed that classical gravity and, therefore, classical spacetime had already come into being.}

\section{Introduction}
Time is running out, but the problem of time in quantum gravity continues to attract attention, as one can judge by recent papers \cite{KP,MM,Rotondo,CC}. This problem arose once the Wheeler -- DeWitt quantum geometrodynamics \cite{DeWitt} was formulated. At first, it caused strong criticism of the Wheeler -- DeWitt approach (see, for example, \cite{Isham,Peres1}). Many efforts were made to find a solution of the problem of time. One possible way to solve this problem is to suggest that some constituent of the Universe (for example, a scalar field or dust matter) could play the role of a clock for the rest of the Universe. Later, however, the other idea was put forward that time had not existed in the Very Early Universe, when it had been of the Planckian size, and time had appeared only in the semiclassical stage of the Universe evolution. This idea is based on the notice that, like there is no trajectories in quantum mechanics, there must be no spacetime (and, therefore, no time itself) in quantum gravity. This point of view is advocated by such well known physicists as Claus Kiefer \cite{Kiefer1} and Carlo Rovelli \cite{Rovelli}. It creates immediately other problems. How should one understand ``the appearance of time''? How could it be treated mathematically? The present paper is devoted to the analysis of the second, ``semiclassical'' approach to the problem of time and comparison of its results with those that can be obtained in the alternative ``extended phase space'' approach.

To the best of our knowledge, one of the first works on semiclassical approximation for gravity was that by Padmanabhan \cite{Padma}, where the gravitational field was treated as a classical object while other fields were quantized (for preceding papers, see \cite{Gerlach,LR}). The study was motivated by difficulties in solving the Wheeler -- DeWitt equation in which all fields are treated as quantum objects. It was supposed that the gravitational field, $g_{\mu\nu}$ is a solution to the Einstein equations in empty space, while the other field describing by some wave function which satisfies some Schr\"odinger equation with a Hamiltonian depending on the background gravitational field $g_{\mu\nu}$. Padmanabhan used a toy quantum mechanical model which mimics gravitation interacting with quantum fields described by the Lagrangian
\begin{equation}
\label{BOA_Lagr}
L=\frac12 M\dot Q^2-MV(Q)+\frac12 m\dot q^2-U(Q,q).
\end{equation}
One can consider this Lagrangian as describing two particles with masses $M$ and $m$. If $M\gg m$, one can use the Born -- Oppenheimer approximation. The position of the heavy particle changes very slowly, and the effect of the light particle on its motion is negligible. The light particle rapidly adjusts to the changes in the position of the heavy particle, but the presence of the latter should be taken into account while describing the light particle. It is believed that the motion of the heavy particle can be described by a classical equation, but the behaviour of the light particle is determined by a temporal Schr\"odinger equation. In this approach, one uses both the semiclassical approximation and an expansion of the action in the parameter $M^{-1}$.

Further, one can make an analogy between the heavy particle and gravity which is supposed to change slowly, and between the light particle and non-gravitational fields. It is not so clear, what is the analogue of the parameter $M^{-1}$. In \cite{Padma}, the gravitational constant $G$ plays the role of $M^{-1}$. In \cite{Singh} the constant $\displaystyle\frac{c^3}{16\pi G}$ that is included into the definition of the gravitational action \cite{LL}, serves as $M$. In other works, the authors consider an expansion in powers of the Planckian mass squared, $m_{Pl}^2=\displaystyle\frac{\hbar c}G$ (see, for example, \cite{KK1,KK2}). In all cases, the limit $M\to\infty$ implies that $G\to 0$. We shall return to this point below. The approach aims at obtaining a temporal Schr\"odinger equation for non-gravitational fields with slowly changing classical gravity in the background. This ``semiclassical program'' was discussed in many works including the monograph \cite{Kiefer1}.

In the essay \cite{KK2} written for the Gravity Research Foundation essay competition and awarded the first prize,
Kiefer and Kr\"amer gave a new strong motivation to this program. They noticed that small effects, like the perihelion precession of Mercury or the Lamb shift, often play the crucial role for the development of physics. Therefore, it would be of great importance to find out small quantum-gravitational effects that can be verified by cosmological observations, in particular, by observation of the anisotropy spectrum of cosmic microwave background radiation (CMB). One can calculate quantum-gravitational corrections to the Schr\"odinger equation for non-gravitational fields in the next order in the expansion in the parameter $M^{-1}$ and make predictions that can be directly compared with observational results. Even more, there is a hope that a comparison of theoretical results with the CMB power spectrum could give researchers the unique opportunity to discriminate between different approaches to quantum gravity.

This goal answers to the greatest dream of cosmologists to turn cosmology into a theory capable of making predictions. In 1999, Stephen Hawking wrote that cosmology was still not a proper science in the sense that it had no predictive power \cite{Hawking1}. Later he stated that cosmology had become a precision science in 2003, with the first results from the WMAP satellite \cite{Hawking2}. Nevertheless, as it was confessed in \cite{KP}, the quantum-gravitational corrections to the CMB power spectrum that can be calculated are too small to be observable in practice.

The applicability of the Born -- Oppenheimer approximation for gravity has been questioned in the recent paper \cite{CC}. The authors emphasized that, when the Born -- Oppenheimer approximation is applied to molecules, it is justified not because of the difference in masses of particles, but because the particles have different dynamical timescales, which are implied when one speaks that the position of the heavy particle changes slowly in time with respect to the light particle. In other words, one needs to have some timescales, and applying the Born --  Oppenheimer approximation to the gravitational field leads us to circularity: one needs time to explain the appearance of time. In \cite{CC} only qualitative analysis is given, while a thorough analysis of calculations can open new details. We intend to make this analysis in the present paper.

As it is used to do in quantum cosmology, we shall consider the closed isotropic model with a scalar field to illustrate basic ideas. In Section 2, we start from the discussion of the WKB approximation for gravity, the Hamilton -- Jacobi equation and constraints. In Section 3, we analyse the ``semiclassical approach'' within the framework of the Wheeler -- DeWitt theory. Then, in Section 4, we show how to apply the same scheme, with necessary modifications, to the extended phase space approach. The latter is known to be gauge-dependent \cite{SSV1,SSV2,SSV3,SSV4}, and results obtained in its framework will be different when using differen gauge conditions. Nevertheless, the Schr\"odinger equations for non-gravitational fields will be the same in the both approaches under the condition on the lapse function $N=1$. On the one hand, it confirms the conclusion made early in our paper \cite{SSV1}, that the Wheeler -- DeWitt approach is a particular case of the extended phase space formulation which corresponds to the special choice of parameterization, gauge conditions and zero eigenvalue of the Hamiltonian. On the other hand, in this particular case we cannot discriminate between the Wheeler -- DeWitt theory and the extended phase space approach, as was hoped in \cite{KK2}. The summary and discussion is given in Section 5. In the present paper, we shall not touch upon related problems, such as the question if the Schr\"odinger equation with quantum-gravitational correction ensures a unitary evolution of quantum states. We hope to address these questions in our future publications.

\section{Gravitational constraints and the WKB approximation}
We shall consider a system of gravitational field with minimal coupling:
\begin{equation}
\label{gen_action}
S=\frac1{4\pi^2}\int\! d^4x\sqrt{-g}\left(\frac{c^3}{16\pi G}R
 +g_{\mu\nu}\partial^{\mu}\phi\partial^{\nu}\phi+m^2\phi^2\right),
\end{equation}
where the coefficient $\displaystyle\frac1{4\pi^2}$ is introduced for convenience, and the coefficient
$\displaystyle\frac{c^3}{16\pi G}$ will be denoted below as $M$.

It is well-known that the Wheeler -- DeWitt quantum geometrodynamics is the application of the Dirac quantization scheme to gravity \cite{Dirac1,Dirac2}. The main Dirac conjecture was that constraints in their operator form must become conditions for a wave function, $\hat T^{\mu}\Psi=0$. The Wheeler -- DeWitt equation is a quantum form of the so called Hamiltonian constraint $T=0$, which is equivalent to $\left(0\atop 0\right)$ Einstein equation, while the three other (momentum) constraints $T^i=0$ correspond to $\left(0\atop i\right)$ Einstein equations, $i=1,2,3$. The gravitational Hamiltonian is a linear combination of constrains, $H=\int d^3x(NT+N_iT^i)$, so that operating on the wave function with the Hamiltonian operator and taking into account the Dirac conjecture, one finds that the result is zero: $\hat H\Psi=\int d^3x(N\hat T\Psi+N_i\hat T^i\Psi)=0$. It leads to the famous problem of time in the Wheeler -- DeWitt quantum geometrodynamics: the wave function does not depend on time.

The Wheeler -- DeWitt equation, in its general form before choosing an operator ordering is:
\begin{equation}
\label{gen_WDW}
\left[-\frac{\hbar^2}M G_{ijkl}\frac{\delta}{\delta\gamma_{ij}}\frac{\delta}{\delta\gamma_{kl}}
 -M\sqrt{\gamma}R^{(3)}
 +H_{(mat)}\left(\phi,\frac{\delta}{\delta\phi}\right)\right]\Psi\left(\gamma_{ij},\phi\right)=0,
\end{equation}
where $\gamma_{ij}$ are space components of the metric tensor, $\gamma$ is the determinant of $\gamma_{ij}$,  $R^{(3)}$ is 3-curvature, $G_{ijkl}$ is the so-called inverse DeWitt supermetric,
\begin{equation}
\label{supermetric}
G_{ijkl}=\frac1{2\sqrt\gamma}\left(\gamma_{ik}\gamma_{jl}+\gamma_{il}\gamma_{jk}-\gamma_{ij}\gamma_{kl}\right).
\end{equation}
Multiplying $G_{ijkl}$ by the lapse function $N$, one gets the inverse metric of configurational space of all positive-definite 3-metrics $\gamma_{ij}$ \cite{HP}. One can say that the DeWitt supermetric is a particular case of the configurational space metric when $N=1$. As a rule, the operator ordering in (\ref{gen_WDW}) is chosen so that the first term in the Wheeler -- DeWitt equation is a Laplacian in the configurational space.

As was said above, we shall use a closed isotropic model. The spacetime interval looks like:
\begin{equation}
\label{interval}
ds^2=N^2(t)dt^2-a^2(t)\left[d\chi^2+\sin^2\chi\left(d\theta^2+\sin^2\theta d\varphi^2\right)\right].
\end{equation}
The scalar field is supposed to be homogeneous. Then, the action for the model is:
\begin{equation}
\label{model_action}
S=\frac12\int\! dt\left(-M\frac{a\dot a^2}N+MNa+\frac{a^3\dot\phi^2}N-Na^3m^2\phi^2\right).
\end{equation}

The scale factor $a$ and the field $\phi$ are physical degrees of freedom of the model, while the lapse function $N$ is the only gauge degree of freedom. As one can see, the metric of configurational space of physical variables is given by
\begin{equation}
\label{Gab}
G_{ab}=\left(
\begin{array}{cc}
-\displaystyle\frac aN&0\\
0&\displaystyle\frac{a^3}N
\end{array}
\right),
\end{equation}
and its inverse is
\begin{equation}
\label{Gab_inv}
G^{ab}=\left(
\begin{array}{cc}
-\displaystyle\frac Na&0\\
0&\displaystyle\frac N{a^3}
\end{array}
\right).
\end{equation}
Let us note that $G_{11}$ component has the minus sign, that is a feature of gravity.

The Wheeler -- DeWitt equation for our model after choosing the operator ordering can be written as:
\begin{equation}
\label{mod_WDW}
\left[-\frac{\hbar^2}{2M}\frac1{\sqrt{-\Gamma}}\frac{\partial}{\partial q^a}
 \left(\sqrt{-\Gamma}\;\Gamma^{ab}\frac{\partial}{\partial q^b}\right)
 -\frac M2a+\frac12 a^3m^2\phi^2\right]\Psi(a,\phi)=0,
\end{equation}
where $\Gamma^{ab}=\displaystyle\frac2N G^{ab}$ is the analogue of the inverse DeWitt supermetric, $q^a=\{a,\phi\}$ are physical degrees of freedom. In its explicit form, Eq.(\ref{mod_WDW}) looks like
\begin{equation}
\label{expl_WDW}
\left[\frac{\hbar^2}{2M}
 \left(\frac1a\frac{\partial^2}{\partial a^2}+\frac1{a^2}\frac{\partial}{\partial a}\right)
 -\frac{\hbar^2}{2}\frac1{a^3}\frac{\partial^2}{\partial\phi^2}
 -\frac M2a+\frac12 a^3m^2\phi^2\right]\Psi(a,\phi)=0,
\end{equation}
It is well-known that, in the WKB approximation, we present the wave function in the form
\begin{equation}
\label{WKB_WF}
\Psi=\exp\left(\frac{iS}{\hbar}\right),
\end{equation}
and then we expand $S$ as
\begin{equation}
\label{WKB_appr}
S=\sigma_0+\frac{\hbar}{i}\sigma_1+\left(\frac{\hbar}{i}\right)^2\sigma_2+\ldots
\end{equation}
Substituting (\ref{WKB_WF}) and (\ref{WKB_appr}) to (\ref{expl_WDW}), in the zero order ${\cal O}(\hbar^0)$ one gets the Hamilton -- Jacobi equation for gravitational and scalar fields:
\begin{equation}
\label{HJE}
\frac1{2Ma}\left(\frac{\partial\sigma_0}{\partial a}\right)^2+\frac12 Ma
 -\frac1{2a^3}\left(\frac{\partial\sigma_0}{\partial\phi}\right)^2-\frac12 a^3m^2\phi^2=0.
\end{equation}
Certainly, if one considers the gravitational action only, Eq. (\ref{HJE}) would be reduced to
\begin{equation}
\label{HJE_grav}
\frac1{2Ma}\left(\frac{\partial\sigma_0}{\partial a}\right)^2+\frac12 Ma=0
\end{equation}

In general, the Hamilton -- Jacobi equation has the form
\begin{equation}
\label{gen_HJE}
H\left(a,\phi,\frac{\partial S}{\partial a},\frac{\partial S}{\partial\phi}\right)=-\frac{\partial S}{\partial t}.
\end{equation}
In the Dirac -- Wheeler -- DeWitt approach \cite{Dirac1,Dirac2,DeWitt} a Hamiltonian is expressed in terms of a linear combination of gravitational constraints, therefore, the right hand side of (\ref{gen_HJE}) is equal to zero. For example, Peres \cite{Peres2} argued that since the Hamiltonian of the gravitational field vanishes weakly, the Hamilton -- Jacobi functional $S$ contains no explicit time dependence. However, if one somehow introduces a reference frame, it results in additional terms in the action, which modify constraints of the original Einstein theory. A good example is given by the Kucha\v r -- Torre model of the so-called reference fluid \cite{KT,MM}. Also, in the extended phase space approach we deal with an effective action which is typical for modern quantum field theory and includes gauge-fixing and ghost terms,
\begin{equation}
\label{eff_act}
S_{(eff)}=S_{(grav)}+S_{(mat)}+S_{(gf)}+S_{(ghost)}.
\end{equation}
Applying the variational procedure to this effective action one obtains modified Einstein equations,
\begin{equation}
\label{EEE}
R_{\mu}^{\nu}-\frac 12\delta_{\mu}^{\nu}R-\kappa T_{\mu(mat)}^{\nu}
 =\kappa\left(T_{\mu(obs)}^{\nu}+T_{\mu(ghost)}^{\nu}\right).
\end{equation}
Here, $T_{\mu(obs)}^{\nu}$ and $T_{\mu(ghost)}^{\nu}$ are not true tensors because they depend on chosen gauge conditions, they are obtained by varying $S_{(gf)}$ and $S_{(ghost)}$, respectively.

In particular, the Hamiltonian constraints corresponding to the $\left(0\atop 0\right)$ Einstein equation is also modified. The left-hand sides of the $\left(0\atop\mu\right)$ modified Einstein equations (\ref{EEE}) are constraints of the original theory. In this approach, they are not equal to zero. In Section 4, we shall see that a Hamiltonian in extended phase space is not a linear combination of constraints. Therefore, we cannot use the argumentation given above that the derivative $\displaystyle\frac{\partial S}{\partial t}$ must be zero. In the extended phase space approach one obtains a temporal Schr\"odinger equation and, accordingly, the Hamilton -- Jacobi equation with a non-zero right-hand side in the zero order ${\cal O}(\hbar^0)$ of the WKB approximation.

It was noticed that, instead of the WKB expansion (\ref{WKB_appr}) one can expand $S$ in powers of $M$, namely,
\begin{equation}
\label{BO_appr}
S=MS_0+S_1+\frac1M S_2+\ldots
\end{equation}
and, under some additional conditions which will be discussed below, in the first order ${\cal O}(M)$, one would obtain the Hamilton -- Jacobi equation for a pure gravitational field (\ref{HJE_grav}). This is the Born -- Oppenheimer approximation for gravity.

In the Born -- Oppenheimer approximation, one should take the limit $M\to\infty$. Let us recall that
$M=\displaystyle\frac{c^3}{16\pi G}$, so that this limit corresponds to $G\to 0$. From a formal mathematical point of view, nothing prevents from taking this limit, but the physical sense of this limit is doubtful, since $G\to 0$ means a total absence of gravity.

To clarify the sense of the expansion parameter $M$, Padmanabhan \cite{Padma} suggested to rewrite (\ref{BOA_Lagr}) as
\begin{equation}
\label{BOA_Lagr1}
\frac LM=\frac12 \dot Q^2-V(Q)+\frac1M\left(\frac12 m\dot q^2-U(Q,q)\right).
\end{equation}
In this expression, $M^{-1}$ can be considered as a coupling constant between heavy and light particles, or, in our case, between gravity and matter fields. However, the coupling between gravitation and matter fields is conventionally encoded in the action of matter fields via metric tensor components. In this approach, the expansion in powers of $M$ (\ref{BO_appr}) serves the purpose to separate the matter fields from gravity and exclude a back reaction of the matter fields on gravity, despite the fact that this situation does not seem realistic.

In the next section, we discuss the Born -- Oppenheimer approximation when applying to gravity.

\section{The semiclassical program for the Wheeler -- DeWitt quantum geometrodynamics}
In this section, we follow the method presented in \cite{KS}. We substitute the expansion (\ref{BO_appr}) into the Wheeler -- DeWitt equation (\ref{expl_WDW}).
\begin{eqnarray}
\label{WDW_expand}
 0&=&\frac{i\hbar}{2Ma}\left(M\frac{\partial^2S_0}{\partial a^2}+\frac{\partial^2S_1}{\partial a^2}
  +\frac1M\frac{\partial^2S_2}{\partial a^2}\right)\nonumber\\
 &-&\frac1{2Ma}\left[M^2\left(\frac{\partial S_0}{\partial a}\right)^2
  +\left(\frac{\partial S_1}{\partial a}\right)^2+\frac1{M^2}\left(\frac{\partial S_2}{\partial a}\right)^2
  +2M\frac{\partial S_0}{\partial a}\frac{\partial S_1}{\partial a}
  +2\frac{\partial S_0}{\partial a}\frac{\partial S_2}{\partial a}
  +\frac2M\frac{\partial S_1}{\partial a}\frac{\partial S_2}{\partial a}\right]\nonumber\\
 &+&\frac{i\hbar}{2Ma^2}\left(M\frac{\partial S_0}{\partial a}
  +\frac{\partial S_1}{\partial a}+\frac1M\frac{\partial S_2}{\partial a}\right)-\frac M2a
  -\frac{i\hbar}{2a^3}\left(M\frac{\partial^2S_0}{\partial\phi^2}
  +\frac{\partial^2S_1}{\partial\phi^2}+\frac1M\frac{\partial^2S_2}{\partial\phi^2}\right)\nonumber\\
 &+&\frac1{2a^3}\left[M^2\left(\frac{\partial S_0}{\partial\phi}\right)^2
  +\left(\frac{\partial S_1}{\partial\phi}\right)^2
  +\frac1{M^2}\left(\frac{\partial S_2}{\partial\phi}\right)^2
  +2M\frac{\partial S_0}{\partial\phi}\frac{\partial S_1}{\partial\phi}
  +2\frac{\partial S_0}{\partial\phi}\frac{\partial S_2}{\partial\phi}
  +\frac2M\frac{\partial S_1}{\partial\phi}\frac{\partial S_2}{\partial\phi}\right]\nonumber\\
 &+&\frac12a^3m^2\phi^2
\end{eqnarray}

\subsection{The orders $\mathbf{\mathcal O(M^2)}$ and $\mathbf{\mathcal O(M)}$}

\noindent\\ In the highest order, ${\cal O}(M^2)$, one has
\begin{equation}
\label{M2}
\frac{\partial S_0}{\partial\phi}=0.
\end{equation}
This equation turns out to be important for the whole approach. Thanks to (\ref{M2}), in the first order
${\cal O}(M)$, one gets the equation which is equivalent to (\ref{HJE_grav}):
\begin{equation}
\label{M1}
\left(\frac{\partial S_0}{\partial a}\right)^2+a^2=0
\end{equation}
(one should remember that $\sigma_0=MS_0$). In other words, one obtains the Hamilton -- Jacobi equation for pure gravity, and so separates gravity from matter fields. It worth noting that Eq. (\ref{M2}) is a consequence of quadratic dependence of the scalar field Hamiltonian on generalized momenta, that gives the second derivative with respect to $\phi$ in (\ref{expl_WDW}), and, in its turn, leads to (\ref{M2}). One can refuse the requirement of homogeneity of the scalar field and present its inhomogeneous part as
\begin{equation}
\label{inhomo}
\delta\phi=\sum_{n,l,m}f_{nlm}(t)\Phi_{nlm}(\chi,\theta,\varphi),
\end{equation}
where $\Phi_{nlm}(\chi,\theta,\varphi)$ are eigenfunctions of Laplacian in a positive curvature space (a similar decomposition was done in \cite{KK1}). One can introduce momenta conjugate to variables $f_{nlm}$ and so come to a Hamiltonian of the scalar field which is quadratic in momenta. This way was chosen in \cite{KK1}. On the other hand, one can go from coordinate representation to the so-called holomorphic representation by expressing operators of generalized coordinates and momenta in terms of creation and annihilation operators. This case has been considered in our recent paper \cite{AKSh1}. In this case, the analogue of Eq. (\ref{WDW_expand}) will not include any terms of the order
${\cal O}(M^2)$, and one cannot obtain an equation like (\ref{M2}). To derive the Hamilton -- Jacobi equation (\ref{M1}), one needs to impose some additional requirements, for example, that $S_0$ does not depend on field variables, as was accepted in \cite{MM}. Obviously, this requirement is equivalent to (\ref{M2}) which cannot be gained automatically. Strictly speaking, this indicates the limitations of the method.

\subsection{The order $\mathbf{\mathcal O(M^0)}$}

\noindent\\ The next order, ${\cal O}(M^0)$, yields:
\begin{equation}
\label{M0}
0=\frac{i\hbar}{2a}\frac{\partial^2S_0}{\partial a^2}
 -\frac1a\frac{\partial S_0}{\partial a}\frac{\partial S_1}{\partial a}
 +\frac{i\hbar}{2a^2}\frac{\partial S_0}{\partial a}
 -\frac{i\hbar}{2a^3}\frac{\partial ^2S_1}{\partial\phi^2}
 +\frac1{2a^3}\left(\frac{\partial S_1}{\partial\phi}\right)^2
 +\frac12 a^3m^2\phi^2.
\end{equation}
Let us introduce the operator
\begin{equation}
\label{H_mat}
H_m=-\frac{\hbar^2}{2a^3}\frac{\partial^2}{\partial\phi^2}
 +\frac12 a^3m^2\phi^2
\end{equation}
and the function
\begin{equation}
\label{chi_fun}
\chi(a,\phi)=D(a)\exp\left(\frac i{\hbar}S_1\right),
\end{equation}
where $D(a)$ is an unknown function. The result of operating on the function $\chi$ with (\ref{H_mat}) is:
\begin{equation}
\label{H_mat_act}
H_m\chi=-\frac{\hbar^2}{2a^3}\frac{\partial^2\chi}{\partial\phi^2}
  +\frac12 a^3m^2\phi^2\chi
 =-\frac{i\hbar}{2a^3}\frac{\partial ^2S_1}{\partial\phi^2}\chi
 +\frac1{2a^3}\left(\frac{\partial S_1}{\partial\phi}\right)^2\chi
 +\frac12 a^3m^2\phi^2\chi.
\end{equation}
Multiplying (\ref{M0}) by $\chi$, one can see that the three last terms in Eq. (\ref{M0}) could be replaced by
$H_m\chi$:
\begin{equation}
\label{M0_Hm}
-\frac{i\hbar}{2a}\frac{\partial^2S_0}{\partial a^2}\chi
 +\frac1a\frac{\partial S_0}{\partial a}\frac{\partial S_1}{\partial a}\chi
 -\frac{i\hbar}{2a^2}\frac{\partial S_0}{\partial a}\chi
 =H_m\chi.
\end{equation}
We would like this equation to have the form of a temporal Schr\"odinger equation with some time variable $\tau$:
\begin{equation}
\label{temp_SE}
i\hbar\frac{\partial\chi}{\partial\tau}=H_m\chi.
\end{equation}

The time variable $\tau$ can be introduced as following. Let us consider a mechanical system with generalized coordinates $q^a$ that satisfy classical motion equations. The state of the system at a given time moment is described by a point in its configurational space. The coordinates $q^a$ change with time, while the point moving along a classical trajectory in the configurational space. The time variable can be chosen as a parameter along the trajectory. Further, a time derivative of an arbitrary function $\chi$ can be written down as
$\displaystyle\frac{\partial\chi}{\partial\tau}=\frac{\partial\chi}{\partial q^a}\frac{\partial q^a}{\partial\tau}$, and, since
$\displaystyle\frac{\partial q^a}{\partial\tau}=G^{ab}p_b=G^{ab}\frac{\partial S}{\partial q^b}$, it finally yields
\begin{equation}
\label{time_oper_gen}
\frac{\partial\chi}{\partial\tau}=G^{ab}\frac{\partial S}{\partial q^b}\frac{\partial\chi}{\partial q^a}.
\end{equation}
The operator in the left-hand side of (\ref{time_oper_gen}) can be presented as
\begin{equation}
\label{time_oper_gen1}
G^{ab}\frac{\partial S}{\partial q^b}\frac{\partial\chi}{\partial q^a}=({\bf p},\nabla)\chi.
\end{equation}
It implies that, in fact, the result of operating is a projection of gradient of $\chi$ on the direction of a tangent vector to a classical trajectory.

In our model, the only generalized coordinate is $a$. Taking into account (\ref{Gab_inv}) and putting $N=1$, the operator in the left-hand side of (\ref{temp_SE}) is
\begin{equation}
\label{time_oper}
i\hbar\frac{\partial\chi}{\partial\tau}=-\frac{i\hbar}a\frac{\partial S_0}{\partial a}\frac{\partial\chi}{\partial a}.
\end{equation}

Operating on (\ref{chi_fun}) with the operator in the right-hand side of (\ref{time_oper}), we have:
\begin{equation}
\label{time_oper_chi}
-\frac{i\hbar}a\frac{\partial S_0}{\partial a}\frac{\partial\chi}{\partial a}
 =-\frac{i\hbar}a\frac{\partial S_0}{\partial a}\frac1D\frac{dD}{da}\chi
  +\frac1a\frac{\partial S_0}{\partial a}\frac{\partial S_1}{\partial a}\chi.
\end{equation}
Equating the left-hand side of (\ref{M0_Hm}) to the right-hand side (\ref{time_oper_chi}), one obtains an equation for $D(a)$, which can be solved taking into account the solution of (\ref{M1}) and yields:
\begin{equation}
\label{D_a}
D(a)=a;\quad
\chi(a,\phi)=a\exp\left(\frac i{\hbar}S_1\right).
\end{equation}

Therefore. with this choice of $\chi$ one would come to the Schr\"odinger equation (\ref{temp_SE}). On the other hand, it should be noted that we need classical trajectories to define the time operator. In other words, a classical spacetime {\it must exist} for one to be able to introduce the time operator and, respectively, the very Schr\"odinger equation. However, it explains by no means {\it how} time (or spacetime) has come into being. Let us remind that in \cite{Kiefer2} Kiefer wrote:
\begin{quote}
``\ldots it is obvious that the emergence of the usual notion of spacetime within quantum cosmology needs an explanation. This is done in two steps: Firstly, a semiclassical approximation to quantum gravity must be performed\ldots This leads to the recovery of an approximate Schr\"odinger equation of non-gravitational fields with respect to the semiclassical background. Secondly, the emergence of classical properties must be explained [by decoherence]\ldots''
\end{quote}
Without doubt, we would like to have an explanation how spacetime has emerged. But the appearance of time is not a process in time. That is the difficulty. We need new ideas, new approaches. It is not enough to perform the semiclassical approximation to say: ``Yes, now we understand how spacetime has emerged from a timeless physical continuum''. Again, we can repeat: one needs spacetime to explain the appearance of spacetime. It is worth noting that the described above method can in no way solve the well-known problem of ``time arrow'' (irreversibility of time).

In \cite{KS}, an analogy between the expansion of the Klein -- Gordon -- Fock equation with respect to the speed of light and the expansion of the Wheeler -- DeWitt equation with respect to the parameter $M=\displaystyle\frac{c^3}{16\pi G}$ is suggested. However, the Klein -- Gordon -- Fock equation is a fundamental relativistic equation, and the Dirac equation for half spin particles can be considered as its consequence. Though many scientists believe that the Wheeler -- DeWitt equation is also fundamental, it has never been verified by observational or experimental data. It is thought that the Wheeler -- DeWitt equation expresses gauge invariance of quantum gravity, however, it has been argued that it is just a particular case of the gauge-dependent Schr\"odinger equation obtained in the framework of the extended pase space approach \cite{Shest1}. Moreover, it is well-known that the expansion in powers of $\displaystyle\frac1c$ gives a non-relativistic limit of small velocities. Therefore, it is to be expected that one obtains the Schr\"odinger equation with relativistic corrections from the fundamental Klein -- Gordon -- Fock equation. On the contrary, the physical sense of the expansion in $\displaystyle\frac1G$ remains obscure.

\subsection{The order $\mathbf{\mathcal O(M^{-1})}$}

\noindent\\ Now we turn to the next order, ${\cal O}(M^{-1})$. We have the equation
\begin{equation}
\label{M-1}
0=\frac{i\hbar}{2a}\frac{\partial^2S_1}{\partial a^2}
  -\frac1{2a}\left[\left(\frac{\partial S_1}{\partial a}\right)^2
   +2\frac{\partial S_0}{\partial a}\frac{\partial S_2}{\partial a}\right]
  +\frac{i\hbar}{2a^2}\frac{\partial S_1}{\partial a}
  -\frac{i\hbar}{2a^3}\frac{\partial^2S_2}{\partial\phi^2}
  +\frac1{a^3}\frac{\partial S_1}{\partial\phi}\frac{\partial S_2}{\partial\phi}.
\end{equation}

Substituting $S_1$ from (\ref{D_a}) one rewrites (\ref{M-1}) as
\begin{equation}
\label{M-1-chi}
\frac1a\frac{\partial S_0}{\partial a}\frac{\partial S_2}{\partial a}-\frac{\hbar^2}{2a^3}
 =\frac{\hbar^2}{2a\chi}\frac{\partial^2\chi}{\partial a^2}
  -\frac{\hbar^2}{2\chi a^2}\frac{\partial\chi}{\partial a}
  -\frac{i\hbar}{\chi a^3}\frac{\partial\chi}{\partial\phi}\frac{\partial S_2}{\partial\phi}
  -\frac{i\hbar}{2a^3}\frac{\partial^2S_2}{\partial\phi^2}.
\end{equation}
Further, let us present $S_2(a,\phi)=\sigma_2(a)+\eta(a,\phi)$, where $\sigma_2(a)$ is a solution to the equation
\begin{equation}
\label{h2}
\frac1a\frac{\partial S_0}{\partial a}\frac{\partial\sigma_2}{\partial a}-\frac{\hbar^2}{2a^3}=0.
\end{equation}
The equation is obtained in the order ${\cal O}(\hbar^2)$ of the WKB expansion (\ref{WKB_appr}). Then,
\begin{equation}
\label{M-1-eta}
\frac1a\frac{\partial S_0}{\partial a}\frac{\partial\eta}{\partial a}
 =\frac{\hbar^2}{2a\chi}\frac{\partial^2\chi}{\partial a^2}
  -\frac{\hbar^2}{2\chi a^2}\frac{\partial\chi}{\partial a}
  -\frac{i\hbar}{\chi a^3}\frac{\partial\chi}{\partial\phi}\frac{\partial\eta}{\partial\phi}
  -\frac{i\hbar}{2a^3}\frac{\partial^2\eta}{\partial\phi^2},
\end{equation}
and we introduce a new function
\begin{equation}
\label{xi-fun}
\xi=\chi\exp\left(\frac{i\eta}{\hbar M}\right).
\end{equation}
After some calculations, using Eq. (\ref{temp_SE}) for $\chi$ and expressing derivatives of $\chi$ through derivatives of $\xi$, one comes to the following equation:
\begin{equation}
\label{M-1-xi}
i\hbar\frac{\partial\xi}{\partial\tau}
 =H_m\xi
 +\frac{\hbar^2}{2M\chi}\left(\frac1a\frac{\partial^2\chi}{\partial a^2}
  -\frac1{a^2}\frac{\partial\chi}{\partial a}\right)\xi.
\end{equation}
This is an analog of Eq. (34) in \cite{KS}. Taking into account that
\begin{equation}
\label{der-chi-xi}
\frac{\partial\chi}{\partial a}
 =\frac{\partial\xi}{\partial a}\frac{\chi}{\xi}+{\cal O}\left(\frac1M\right);
\quad
\frac{\partial^2\chi}{\partial a^2}
 =\frac{\partial^2\xi}{\partial a^2}\frac{\chi}{\xi}+{\cal O}\left(\frac1M\right),
\end{equation}
we finally obtain:
\begin{equation}
\label{xi-fin}
i\hbar\frac{\partial\xi}{\partial\tau}
 =H_m\xi
 +\frac{\hbar^2}{2M}\left(\frac1a\frac{\partial^2\xi}{\partial a^2}
  -\frac1{a^2}\frac{\partial\xi}{\partial a}\right).
\end{equation}

This is the required Schr\"odinger equation with quantum gravitational corrections.

\section{The extended phase space approach}
Most attempts of quantization of gravity imply application of methods elaborated for non-gravitational fields. As a rule, the features of gravity are not taken into account. The picture that is accepted in a conventional quantum field theory is that there are free non-interacting particles in initial and final states, which affect each other in the interaction region. The initial and final states with free particles are known as asymptotic states. In the path integral technique, the presence of asymptotic states requires imposing asymptotic boundary conditions which ensure gauge invariance of the theory.

This picture does not correspond to gravity. Gravitating systems do not have asymptotic states, except asymptotically-flat spacetimes. Even in so simple model, as that of an isotropic universe, asymptotic states do not exist. This means that proofs of gauge invariance, like the Fradkin -- Vilkovisky theorem \cite{Henneaux}, are not valid in the case of gravity. Gauge invariance of quantum gravity is assumed by many authors, but it has not been demonstrated at a strong mathematical level. Our own study raises doubts that quantum gravity can be constructed in a gauge-invariant way \cite{SSV1,SSV2,SSV3,SSV4,Shest1}.

In the extended phase space approach asymptotic boundary conditions are rejected as inappropriate to the physical situation. As was said in Section 2, we use the effective action (\ref{eff_act}). For our model it reads:
\begin{equation}
\label{model_eff_act}
S_{(eff)}
 =\int\! dt\left[\frac12\left(-M\frac{a\dot a^2}N+MNa+\frac{a^3\dot\phi^2}N-Na^3m^2\phi^2\right)
  +\pi\left(\dot N-\frac{df}{da}\dot a\right)+\dot{\bar\theta}N\bar\theta\right].
\end{equation}
Here $\theta$, $\bar\theta$ are the Faddeev -- Popov ghosts; the gauge condition $N=f(a)+k$, $k=\rm{const}$ is used in a differential form, $\dot N=\displaystyle\frac{df}{da}\dot a$. It introduces the missing velocity $\dot N$ into the effective Lagrangian that enables one to construct the Hamiltonian in extended phase space according to the usual rule,
$H=p\dot q-L$, so one gets:
\begin{equation}
\label{EPS_Ham}
H=-\frac12\frac N{Ma}\left(p_a+\pi\frac{df}{da}\right)^2
  +\frac12\frac N{a^3}p_{\phi}^2
  -\frac12 MNa +\frac12 Na^3m^2\phi^2+\frac1N\bar{\cal P}{\cal P},
\end{equation}
where $p_a$, $p_{\phi}$ are the momenta conjugate to $a$ and $\phi$, correspondingly, $\pi$ is the momentum conjugate to $N$ and $\bar{\cal P}$, $\cal P$ are ghost momenta. In the Dirac approach, a Hamiltonian must be a linear combination of constraints,
\begin{equation}
\label{Dir_Ham}
H_D=-\frac N2\left(\frac1{Ma}p_a^2-\frac1{a^3}p_{\phi}^2+Ma-a^3m^2\phi^2\right).
\end{equation}
In the extended phase space approach, the constraint is modified and has the form:
\begin{equation}
\label{Ham_constr}
\frac1{2Ma}\left(p+\pi\frac{df}{da}\right)^2-\frac1{2a^3}p_{\phi}^2
 +\frac12 Ma-\frac12a^3m^2\phi^2+\frac1{N^2}\bar{\cal P}{\cal P}-\dot\pi=0.
\end{equation}
One can see that the Hamiltonian in extended phase cannot be presented as (\ref{Ham_constr}) multiplied by $N$.

Following the method elaborated by Feynman \cite{Feynman} and generalised by Cheng \cite{Cheng}, we derive the Schr\"odinger equation for our model from the path integral with the effective action (\ref{model_eff_act}):
\begin{equation}
\label{Sch_eq_full}
i\frac{\partial\Psi}{\partial t}
 =-\frac1{2M\mu}\frac{\partial}{\partial Q^{\alpha}}
   \left(\mu G^{\alpha\beta}\frac{\partial\Psi}{\partial Q^{\beta}}\right)
  -\frac M2 Na\Psi+\frac12 Na^3m^2\phi^3\Psi,
\end{equation}
where $\mu$ is the measure in the path integral, for our model $\mu=\displaystyle\frac{a^2}N$; $Q^{\alpha}=(N,a,\phi)$;
\begin{equation}
\label{Galbe_inv}
G^{\alpha\beta}=\left(
\begin{array}{ccc}
-\displaystyle\frac Na\left(\frac{df}{da}\right)^2&-\displaystyle\frac Na\frac{df}{da}&0\\
-\displaystyle\frac Na\frac{df}{da}&-\displaystyle\frac Na&0\\
0&0&\displaystyle\frac N{a^3}
\end{array}
\right);
\end{equation}
the wave function $\Psi(N,a,\phi,\theta,\bar\theta;t)$ is defined on extended configurational space which includes physical as well as gauge and ghost variables. The general solution to the equation (\ref{Sch_eq_full}) is:
\begin{equation}
\label{GS}
\Psi(N,a,\phi,\theta,\bar\theta;t)
 =\int\Psi_k(a,\phi,t)\,\delta(N-f(a)-k)\,(\bar\theta+i\theta)\,dk.
\end{equation}

Our main goal is the Schr\"odinger equation for the physical part of the wave function $\Psi_k(a,\phi,t)$, whose explicit form for our model reads
\begin{equation}
\label{Sch_phys}
i\hbar\frac{\partial\Psi}{\partial t}
 =\frac{\hbar^2}{2M}\frac{f(a)}{a^2}\left(a\frac{\partial^2\Psi}{\partial a^2}+\frac{\partial\Psi}{\partial a}\right)
  -\frac M2 f(a)a\Psi-\frac{\hbar^2}2\frac{f(a)}{a^3}\frac{\partial^2\Psi}{\partial\phi^2}
  +\frac12f(a)a^3m^2\phi^2\Psi.
\end{equation}
One can see that if $\displaystyle\frac{\partial\Psi}{\partial t}=0$ and $f(a)=1$, this equation would be reduced to (\ref{expl_WDW}).

We now reproduce the semiclassical program, substituting the expansion (\ref{BO_appr}) into (\ref{Sch_phys}):
\begin{eqnarray}
\label{Sch_expand}
-M\frac{\partial S_0}{\partial t}\!
&\!-\!&\!\frac{\partial S_1}{\partial t}-\frac1M\frac{\partial S_2}{\partial t}\nonumber\\
&=&\frac{i\hbar}{2M}\frac{f(a)}a\left(M\frac{\partial^2S_0}{\partial a^2}+\frac{\partial^2S_1}{\partial a^2}
 +\frac1M\frac{\partial^2S_2}{\partial a^2}\right)\nonumber\\
&-&\frac1{2M}\frac{f(a)}a\left[M^2\left(\frac{\partial S_0}{\partial a}\right)^2
 +\left(\frac{\partial S_1}{\partial a}\right)^2+\frac1{M^2}\left(\frac{\partial S_2}{\partial a}\right)^2
 +2M\frac{\partial S_0}{\partial a}\frac{\partial S_1}{\partial a}
 +2\frac{\partial S_0}{\partial a}\frac{\partial S_2}{\partial a}
 +\frac2M\frac{\partial S_1}{\partial a}\frac{\partial S_2}{\partial a}\right]\nonumber\\
&+&\frac{i\hbar}{2M}\frac{f(a)}{a^2}\left(M\frac{\partial S_0}{\partial a}
 +\frac{\partial S_1}{\partial a}+\frac1M\frac{\partial S_2}{\partial a}\right)-\frac M2 f(a)a
 -\frac{i\hbar}2\frac{f(a)}{a^3}\left(M\frac{\partial^2S_0}{\partial\phi^2}
 +\frac{\partial^2S_1}{\partial\phi^2}+\frac1M\frac{\partial^2S_2}{\partial\phi^2}\right)\nonumber\\
&+&\frac12\frac{f(a)}{a^3}\left[M^2\left(\frac{\partial S_0}{\partial\phi}\right)^2
 +\left(\frac{\partial S_1}{\partial\phi}\right)^2
 +\frac1{M^2}\left(\frac{\partial S_2}{\partial\phi}\right)^2
 +2M\frac{\partial S_0}{\partial\phi}\frac{\partial S_1}{\partial\phi}
 +2\frac{\partial S_0}{\partial\phi}\frac{\partial S_2}{\partial\phi}
 +\frac2M\frac{\partial S_1}{\partial\phi}\frac{\partial S_2}{\partial\phi}\right]\nonumber\\
&+&\frac12 f(a)a^3m^2\phi^2.
\end{eqnarray}

\subsection{The orders $\mathbf{\mathcal O(M^2)}$ and $\mathbf{\mathcal O(M)}$}

\noindent\\ The order ${\cal O}(M^2)$ again yields Eq. (\ref{M2}), while in the order ${\cal O}(M)$ one comes to the expression:
\begin{equation}
\label{M1_EPS}
-\frac{\partial S_0}{\partial t}
 =-\frac12\frac{f(a)}a\left(\frac{\partial S_0}{\partial a}\right)^2-\frac12 f(a)a=0.
\end{equation}

To be in agreement with the classical Einstein theory and obtain the Hamilton -- Jacobi equation, one should put
$\displaystyle\frac{\partial S_0}{\partial t}=0$. Then, the gauge-fixing function $f(a)$ can be cancelled out and one restores Eq. (\ref{M1}). Therefore, in these two orders we have no new results in comparison with the Wheeler -- DeWitt approach.

\subsection{The order $\mathbf{\mathcal O(M^0)}$}

\noindent\\ In the order ${\cal O}(M^0)$ we have:
\begin{equation}
\label{M0_EPS}
-\frac{\partial S_1}{\partial t}
 =\frac{i\hbar}2\frac{f(a)}a\frac{\partial^2S_0}{\partial a^2}
  -\frac{f(a)}a\frac{\partial S_0}{\partial a}\frac{\partial S_1}{\partial a}
  +\frac{i\hbar}2\frac{f(a)}{a^2}\frac{\partial S_0}{\partial a}
  -\frac{i\hbar}2\frac{f(a)}{a^3}\frac{\partial ^2S_1}{\partial\phi^2}
  +\frac12\frac{f(a)}{a^3}\left(\frac{\partial S_1}{\partial\phi}\right)^2
  +\frac12 f(a)a^3m^2\phi^2.
\end{equation}
To obtain a Schr\"odinger equation for matter fields similar to (\ref{temp_SE}), we do not need to construct an operator like (\ref{time_oper}), since our equations have already included time variable $t$. Instead of (\ref{H_mat}) we introduce the operator
\begin{equation}
\label{H_mat_EPS}
H_{mat}=-\frac{\hbar^2}2\frac{f(a)}{a^3}\frac{\partial^2}{\partial\phi^2}
 +\frac12 f(a)a^3m^2\phi^2.
\end{equation}
Again, after multiplying (\ref{M0_EPS}) by $\chi$ (\ref{chi_fun}) one can see that the three last terms in (\ref{M0_EPS}) could be replaced by $H_{mat}\chi$. Therefore, to get the Schr\"odinger equation
\begin{equation}
\label{temp_SE_EPS}
i\hbar\frac{\partial\chi}{\partial t}=H_{mat}\chi
\end{equation}
one needs that the first three terms in the right-hand side of (\ref{M0_EPS}) would be equal to zero:
\begin{equation}
\label{S1}
\frac{i\hbar}2\frac{\partial^2S_0}{\partial a^2}
 -\frac{\partial S_0}{\partial a}\frac{\partial S_1}{\partial a}
 +\frac{i\hbar}{2a}\frac{\partial S_0}{\partial a}=0.
\end{equation}
One can consider this as an equation for $S_1$. However, we now do not have an equation for $D(a)$. One can put $D(a)=a$ for correspondence with what one obtains in the Wheeler -- DeWitt approach, but one can also choose $D(a)=1$.

\subsection{The order $\mathbf{\mathcal O(M^{-1})}$}

\noindent\\ In this order, one obtains the equation:
\begin{equation}
\label{M-1_EPS}
-\frac{\partial S_2}{\partial t}
 =i\hbar\frac{f(a)}{2a}\frac{\partial^2S_1}{\partial a^2}
  -\frac{f(a)}{2a}\left(\frac{\partial S_1}{\partial a}\right)^2
  -\frac{f(a)}a\frac{\partial S_0}{\partial a}\frac{\partial S_2}{\partial a}
  +i\hbar\frac{f(a)}{2a^2}\frac{\partial S_1}{\partial a}
  -i\hbar\frac{f(a)}{2a^3}\frac{\partial^2S_2}{\partial\phi^2}
  +\frac{f(a)}{a^3}\frac{\partial S_1}{\partial\phi}\frac{\partial S_2}{\partial\phi}.
\end{equation}

Returning to the function (\ref{chi_fun}), we express all derivatives of $S_1$ and substitute them into (\ref{M-1_EPS}):
\begin{eqnarray}
\label{M-1_EPS-1}
-\frac{\partial S_2}{\partial t}
&=&\hbar^2\frac{f(a)}{2a}\left[\frac1{\chi}\frac{\partial^2\chi}{\partial a^2}
  -\frac1D\frac{d^2D}{da^2}+\frac2{D^2}\left(\frac{dD}{da}\right)^2
  -\frac2{D\chi}\frac{dD}{da}\frac{\partial\chi}{\partial a}\right]
 -\frac{f(a)}a\frac{\partial S_0}{\partial a}\frac{\partial S_2}{\partial a}\nonumber\\
&+&\hbar^2\frac{f(a)}{2a^2}\left(\frac1{\chi}\frac{\partial \chi}{\partial a}
  -\frac1D\frac{dD}{da}\right)
 -i\hbar\frac{f(a)}{2a^3}\frac{\partial^2S_2}{\partial\phi^2}
 -i\hbar\frac{f(a)}{\chi a^3}\frac{\partial\chi}{\partial\phi}\frac{\partial S_2}{\partial\phi}.
\end{eqnarray}

To reproduce the results of the previous Section, the solution is sought in the form $S_2(a,\phi)=\sigma_2(a)+\eta(\phi)$, at that, $\sigma_2(a)$ does not depend on $t$, as supposed in the Wheeler -- DeWitt approach. We require that $\sigma_2(a)$ is a solution to the equation
\begin{equation}
\label{sigma2}
0=-\frac1a\frac{\partial S_0}{\partial a}\frac{\partial\sigma_2}{\partial a}
  -\frac{\hbar^2}{2a}\left[\frac1D\frac{d^2D}{da^2}-\frac2{D^2}\left(\frac{dD}{da}\right)^2\right]
  -\frac{\hbar^2}{2Da^2}\frac{dD}{da}.
\end{equation}
When $D(a)=a$, we come to (\ref{h2}).

Let us note that now the function $\eta$ does not depend on $a$. This differs from what we did in Section 3, but this enables us to exclude the term
$-\displaystyle\frac{f(a)}a\frac{\partial S_0}{\partial a}\frac{\partial\eta}{\partial a}$,
which we do not need to construct the time derivative (see (\ref{time_oper})). The equation for $\eta$ is:
\begin{equation}
\label{eta}
-\frac{\partial\eta}{\partial t}
 =\hbar^2\frac{f(a)}{2a}\left(\frac1{\chi}\frac{\partial^2\chi}{\partial a^2}
  -\frac2{D\chi}\frac{dD}{da}\frac{\partial\chi}{\partial a}\right)
 +\hbar^2\frac{f(a)}{2a^2}\frac1{\chi}\frac{\partial\chi}{\partial a}
 -i\hbar\frac{f(a)}{2a^3}\frac{\partial^2\eta}{\partial\phi^2}
 -i\hbar\frac{f(a)}{\chi a^3}\frac{\partial\chi}{\partial\phi}\frac{\partial\eta}{\partial\phi}.
\end{equation}
When $D(a)=a$, this equation is reduced to the following:
\begin{equation}
\label{eta_Da}
-\frac{\partial\eta}{\partial t}
 =\hbar^2\frac{f(a)}{2a\chi}\left(\frac{\partial^2\chi}{\partial a^2}
 -\frac1a\frac{\partial\chi}{\partial a}\right)
 -i\hbar\frac{f(a)}{2a^3}\frac{\partial^2\eta}{\partial\phi^2}
 -i\hbar\frac{f(a)}{a^3\chi}\frac{\partial\chi}{\partial\phi}\frac{\partial\eta}{\partial\phi}.
\end{equation}

Now, as in Section 3.3, we introduce the function (\ref{xi-fun}). Using (\ref{temp_SE_EPS}), one comes to the equation
\begin{equation}
\label{a-la-Kiefer}
i\hbar\frac{\partial\xi}{\partial\tau}
 =H_{mat}\xi
 +\hbar^2\frac{f(a)}{2M}\left(\frac1a\frac{\partial^2\xi}{\partial a^2}
  -\frac1{a^2}\frac{\partial\xi}{\partial a}\right).
\end{equation}
It is similar to (\ref{xi-fin}) and coincides with the latter if $f(a)=1$. Therefore, in this particular case the quantum gravitational corrections are not distinguishable from those obtained in Section 3.3 for the Wheeler -- DeWitt approach.

Let us emphasize that, early in our papers, it has been demonstrated that the Wheeler -- DeWitt equation can be obtained from the Schr\"odinger equation for the physical part of the wave function if one chooses the Arnowitt -- Deser -- Misner (ADM) parametrization and the gauge conditions on the lapse and shift function $N=1$, $N_i=0$, and rejects the time evolution (see \cite{SSV1}). In the present work, we do not need to reject time dependence of the wave function. On the opposite, in the Wheeler -- DeWitt approach, one has to construct a time derivative and restore a temporal Schr\"odinger equation.

If the Wheeler -- DeWitt equation can be obtained from the Schr\"odinger equation as a particular case, the reader can inquire: is the extended phase space approach more general than the Wheeler -- DeWitt one? The answer that can be obtained in the framework of the extended phase approach is yes. The Wheeler -- DeWitt equation is a direct consequence of the Dirac postulate, according to which, after quantization, the constraints become conditions on the wave function. It reduces the Hamiltonian spectrum to the only eigenvalue $E=0$ and leads to the problem of time. The conditions $N=1$, $N_i=0$ were implied yet in the paper by DeWitt \cite{DeWitt} and implicitly introduce a privileged reference frame. There is no proof that the wave function satisfying the Wheeler -- DeWitt equation is gauge invariant. In the extended phase space approach, one gets a physical picture which is explicitly non-invariant and depends on a chosen reference frame, but all reference frames are equal, that corresponds to the spirit of general relativity, where pictures, watching by different observers, are complementary to each other.

However, to reproduce the results for the Wheeler -- DeWitt approach in our model, it is not enough to choose the condition $N=1$. One should also put $D(a)=a$, though it is not compulsory in the extended phase space approach. Actually, to the order ${\cal O}(M^0)$, the wave function is
$\Psi=\exp\left[\displaystyle\frac i{\hbar}(MS_0+S_1)\right]$, but not
$\Psi=D(a)\exp\left[\displaystyle\frac i{\hbar}(MS_0+S_1)\right]$. Multiplication on the function $D(a)$ changes the measure in the inner product, even if the matter part of the wave function is considered. Nevertheless, in the approach presented in Section 3, introducing the function $D(a)$ is necessary to obtain the time operator.

Choosing $D(a)=1$, instead of (\ref{M-1_EPS-1}) one would have the equation:
\begin{equation}
\label{D-1}
-\frac{\partial S_2}{\partial t}
=\hbar^2\frac{f(a)}{2a}\frac1{\chi}\frac{\partial^2\chi}{\partial a^2}
 -\frac{f(a)}a\frac{\partial S_0}{\partial a}\frac{\partial S_2}{\partial a}
 +\hbar^2\frac{f(a)}{2a^2}\frac1{\chi}\frac{\partial \chi}{\partial a}
 -i\hbar\frac{f(a)}{2a^3}\frac{\partial^2S_2}{\partial\phi^2}
 -i\hbar\frac{f(a)}{\chi a^3}\frac{\partial\chi}{\partial\phi}\frac{\partial S_2}{\partial\phi}.
\end{equation}
The second term in the right-hand side of (\ref{D-1}) seems to be redundant for our purpose. We can avoid it putting $S_2(a,\phi)=\eta(\phi)$. Proceeding as in the previous case, one gets the following equation with quantum gravitational corrections:
\begin{equation}
\label{not-Kiefer}
i\hbar\frac{\partial\xi}{\partial\tau}
 =H_{mat}\xi
 +\hbar^2\frac{f(a)}{2M}\left(\frac1a\frac{\partial^2\xi}{\partial a^2}
  +\frac1{a^2}\frac{\partial\xi}{\partial a}\right).
\end{equation}
The equation would include the additional term of the order ${\cal O}(M^0)$ (which is the second term in the right-hand side of (\ref{D-1})) if we did not accept that $S_2$ does not depend on $a$. It differs from (\ref{a-la-Kiefer}) only by the sign ``+'' before the last term. In principle, there exists a hope that it may be enough to discriminate between different approaches if a desirable accuracy of observations is achieved.

\section{Discussion and conclusions}
We have seen in the pervious sections that the task of obtaining a temporal Schr\"odinger equation for matter fields with quantum gravitational corrections can be solved in different ways in various approaches. And even in the framework of the extended phase space approach one can get different forms of this equation. A final form of this equation depends on additional assumptions which should be made inevitably when tackling the problem. On the other hand, adopting some assumptions one can obtain the same form of the equation as the one in another approach.

In this paper, we have analysed mainly the approach applied by Kiefer and his collaborators in \cite{KS,KK1,KP} and other papers and questioned if these result can be reproduced in the framework of the extended phase space approach. We intentionally illustrated the main steps with the closed isotropic model to make all calculations transparent.

To give another example, let us discuss briefly the approach used by Montani and his collaborators \cite{MM,GMMN}. They also have the goal to get a Schr\"odinger equation for matter fields, but not to explain the appearance of time in quantum gravity. In their approach, time is appeared thanks to introducing the Kucha\v r -- Torre reference fluid. They present the wave function in the form (\ref{WKB_WF}), (\ref{BO_appr}), where
\begin{equation}
\label{S_sep}
S_0(a)=\sigma_0(a);\quad
S_1(a,\phi)=\sigma_1(a)+Q_1(a,\phi);\quad
S_2(a,\phi)=\sigma_2(a)+Q_2(a,\phi).
\end{equation}
The authors have done a number of assumptions, in particular, that the gravitational part of the wave function must satisfy constraints for pure gravity. Therefore, many terms in the equations are disappeared, and we have in the orders ${\cal O}(M^0)$ and ${\cal O}(M^{-1})$, correspondingly:
\begin{equation}
\label{MM-M0}
-\frac{\partial Q_1}{\partial t}
 =-\frac{f(a)}a\frac{\partial\sigma_0}{\partial a}\frac{\partial Q_1}{\partial a}
  -\frac{i\hbar}2\frac{f(a)}{a^3}\frac{\partial^2Q_1}{\partial\phi^2}
  +\frac12\frac{f(a)}{a^3}\left(\frac{\partial Q_1}{\partial\phi}\right)^2+\frac12f(a)a^3m^2\phi^2
\end{equation}
and
\begin{equation}
\label{MM-M-1}
-\frac{\partial Q_2}{\partial t}
 =i\hbar\frac{f(a)}{2a}\frac{\partial^2Q_1}{\partial a^2}
  -\frac{f(a)}{2a}\left[2\frac{\partial\sigma_1}{\partial a}\frac{\partial Q_1}{\partial a}
   +\left(\frac{\partial Q_1}{\partial a}\right)^2
   +2\frac{\partial\sigma_0}{\partial a}\frac{\partial Q_2}{\partial a}\right]
  +i\hbar\frac{f(a)}{2a^2}\frac{\partial Q_1}{\partial a}
  -i\hbar\frac{f(a)}{2a^3}\frac{\partial^2Q_2}{\partial\phi^2}
  +\frac{f(a)}{a^3}\frac{\partial Q_1}{\partial\phi}\frac{\partial Q_2}{\partial\phi}.
\end{equation}

Further, the function $\xi$ is introduced (compare with (\ref{xi-fun-MM})):
\begin{equation}
\label{xi-fun-MM}
\xi=\chi\exp\left(\frac{iQ_2}{\hbar M}\right)
   =\exp\left[\frac i{\hbar}\left(Q_1+\frac1M Q_2\right)\right].
\end{equation}

Eq. (\ref{MM-M-1}) is divided by $M$ and added to Eq. (\ref{MM-M0}), after which one comes to the following equation for $\xi$:
\begin{equation}
\label{xi-MM-1}
i\hbar\frac{\partial\xi}{\partial t}
 =i\hbar\frac{f(a)}a\frac{\partial\sigma_0}{\partial a}\frac{\partial\xi}{\partial a}
  -\hbar^2\frac{f(a)}{2a^3}\frac{\partial^2\xi}{\partial\phi^2}
  +\frac12 f(a)a^3m^2\phi^2\xi
  +\hbar^2\frac{f(a)}{2Ma\chi}\frac{\partial^2\chi}{\partial a^2}\xi
  +\hbar^2\frac{f(a)}{2Ma^2\chi}\frac{\partial\chi}{\partial a}\xi
  +i\hbar\frac{f(a)}{Ma\chi}\frac{\partial\sigma_1}{\partial a}\frac{\partial\chi}{\partial a}\xi.
\end{equation}

According to another assumption of the authors, the derivatives of the matter part of the wave function with respect to slowly changing gravitational variables are small:
\begin{equation}
\label{deriv_MM}
\frac{\partial\chi}{\partial a}={\cal O}\left(\frac1M\right).
\end{equation}
Then, the last three terms in (\ref{xi-MM-1}) are of the order ${\cal O}(M^{-2})$. The only quantum correction is the first term in the right-hand side. So, the final form of the equation in this approach is:
\begin{equation}
\label{xi-MM-fin}
i\hbar\frac{\partial\xi}{\partial t}
 =H_{mat}\xi+i\hbar\frac{f(a)}a\frac{\partial\sigma_0}{\partial a}\frac{\partial\xi}{\partial a}.
\end{equation}

The term
$i\hbar\displaystyle\frac{f(a)}a\frac{\partial\sigma_0}{\partial a}\frac{\partial\xi}{\partial a}$
is considered in \cite{MM} as a quantum gravitational correction. The authors believe that this correction is Hermitian and does not break down unitary evolution, in contrast to the corrections in the equations (\ref{xi-fin}), (\ref{a-la-Kiefer}) and (\ref{not-Kiefer}), which are not Hermitian, as one can check. The authors consider this as an advantage of their approach. However, the term is Hermitian if
\begin{enumerate}
  \item the classical action for gravity $\sigma_0$ is real. It is not true for the closed isotropic model and means that, at least, this approach cannot be applied to the closed model;
  \item the measure in the inner product is trivial. In the extended phase space approach, the measure in the inner product is the same as the measure in the path integral; the latter is determined by the procedure of derivation of the Schr\"odinger equation (\ref{Sch_eq_full}) from the path integral. In other approaches, an inner product in Hilbert space should be defined unambiguously.
\end{enumerate}

We can see that in the three approaches discussed in the present paper, the equations with different quantum gravitational corrections have been obtained, and the corrections are very different. Their form depends on additional assumptions, which are rather arbitrary and not of a fundamental character. On the one hand, it gives some hope to discriminate between different approaches to quantum gravity in the future when we get more accurate data. On the other hand, we need more reasonable principles to derives the sought equation for matter fields with quantum gravitational corrections. And, in any case, none of the approaches can explain how time had appeared in the Early Universe, since it is supposed that classical gravity and, therefore, classical spacetime had already come into being.

\end{document}